\documentclass[12pt,preprint]{iopart}
\usepackage{graphicx}

\begin{document}

\title[FPT statistics of a Brownian motion driven by an exponential temporal drift]{Series solution to the first-passage-time problem of a Brownian motion with an exponential time-dependent drift}

\author{Eugenio Urdapilleta}
\address{Divisi\'on de F\'isica Estad\'istica e Interdisciplinaria \& Instituto Balseiro, Centro At\'omico Bariloche, Av. E. Bustillo Km 9.500, S. C. de Bariloche (8400), R\'io Negro, Argentina}
\ead{urdapile@ib.cnea.gov.ar}
\begin{abstract}
We derive the first-passage-time statistics of a Brownian motion
driven by an exponential time-dependent drift up to a threshold.
This process corresponds to the signal integration in a simple
neuronal model supplemented with an adaptation-like current and
reaching the threshold for the first time represents the condition
for declaring a spike. Based on the backward Fokker-Planck
formulation, we consider the survival probability of this process
in a domain restricted by an absorbent boundary. The solution is
given as an expansion in terms of the intensity of the
time-dependent drift, which results in an infinite set of
recurrence equations. We explicitly obtain the complete solution
by solving each term in the expansion in a recursive scheme. From
the survival probability, we evaluate the first-passage-time
statistics, which itself preserves the series structure. We then
compare theoretical results with data extracted from numerical
simulations of the associated dynamical system, and show that the
analytical description is appropriate whenever the series is
truncated in an adequate order.
\end{abstract}

\maketitle

\section{Introduction}
The statistical analysis of a system is essential when
fluctuations contribute to its dynamics \cite{vankampen,
ricciardi, hanggi2005}. In different areas, beyond the importance
of the statistical description of the system state and its
evolution, the main variable of interest is the time at which this
state reaches a certain region for the first time \cite{redner},
constituting the so-called first-passage-time (FPT) problem. For
example, in a diffusion-controlled reaction a particle performs a
random walk until it makes contact with a reactant or a trap
giving rise to the reaction \cite{redner}. Generally, in a FPT
problem, the system is able to evolve according to a given
dynamics in a confined region, limited by one or more absorbing
boundaries. Even for simple autonomous systems, the FPT problem
can be analytically difficult to solve. For example, for the
Ornstein-Uhlenbeck process with a fixed positive absorbing
boundary, the FPT solution (representing the FPT density function)
is relatively easy to be found in the Laplace domain, but its
inverse transform is not explicitly available \cite{tuckwell}. A
notable exception is the FPT problem for a Brownian motion (Wiener
process), where different methods are easy to be applied to solve
it \cite{vankampen, ricciardi, redner, tuckwell, gerstein1964}. A
greater complexity is found when a time-inhomogeneous process
defines the system dynamics. In this case, analytical methods are
\textit{formally} given within different approaches
\cite{ricciardi, redner, tuckwell, risken, gardiner}, but exact as
well as approximate explicit results are scarce in the literature
and probably difficult to obtain. Among the different ways to
introduce a time inhomogeneity into the system (for example,
temporally varying absorbing boundaries \cite{lindner2005,
tuckwell1984}, time-dependent drift and diffusion coefficients
\cite{molini2011}, etc), we focus on a particular drift
coefficient evolving externally in time. The process analysed here
is driven by an exponential time-dependent drift (this case, in
turn, can be mapped into a variable threshold \cite{lindner2005}),
which naturally arises in neuroscience when modelling adapting
neurons \cite{madison1984, helmchen1996, sah1996, liu2001,
benda2003, benda2010}. In this context, the membrane potential
(state variable) of a perfect integrate-and-fire neuron model is
driven during its subthreshold evolution by an external current,
composed of a constant deterministic current plus fast
fluctuations \cite{ricciardi, tuckwell, gerstein1964, gerstner,
burkitt1}, as well as an intrinsic temporally decaying current
\cite{benda2010, schwalger2010}. This system corresponds exactly
to the case analysed here, and the FPT represents the production
of a spike.\\
\indent The interest in the FPT problem for a Brownian particle in
a time-inhomogeneous setup started in the 1990s, when different
systems driven by periodically modulated drifts were studied
within the context of the stochastic resonance phenomenon
\cite{gammaitoni1998}. In particular, Bulsara \textit{et al}
applied the method of images to a Wiener process driven by a
sinusoidal temporal drift in the presence of an absorbing boundary
\cite{bulsara1994}, a procedure of limited validity
\cite{gitterman1995, bulsara1995, molini2011}. Later, other
threshold processes under analogous conditions were theoretically
analysed with different approximation methods
\cite{gammaitoni1998, bulsara1996, schindler2004, burkitt2}. Even
when appealing, single sinusoidal temporal drifts do not represent
a general case. For arbitrary time-dependent drifts, the simplest
procedures are a quasiadiabatic reduction \cite{choi2002}, a
quasistatic description \cite{urdapilleta2009}, or small amplitude
approximations. However, for rapidly varying arbitrary fields, the
time-dependent structure of the problem cannot be simplified. Up
to our knowledge, the first attempt to include an arbitrary
temporal drift (without spatial dependence) in a general framework
was made in \cite{lindner2004}, where the author proposed a method
to describe the first-order correction to the moments of the FPT
density, in a perturbation scheme.\\
\indent The preceding studies describe \textit{approximately} the
FPT problem of a given continuous stochastic process in a
restricted domain for particular or general temporal drifts. In
general, explicit \textit{exact} results for time-inhomogeneous
systems are infrequent. Notably and as an exception, in
\cite{molini2011} the authors derive the FPT density function of a
Wiener process, in the presence of an absorbing boundary, where
both the drift and the diffusion coefficients are varying
temporally and in proportion to each other. In particular, when
proportionality is satisfied, the Fokker-Planck equation ruling
the evolution of the transition probability between the states at
two times can be time-rescaled in order to resemble the simpler
constant coefficients case, where the exact solution is known.
However, the restriction in the coefficients proportionality
limits its applicability to our case. For the FPT problem we are
interested in, we have previously proposed a series solution in
terms of the intensity of the time-dependent drift
\cite{urdapilleta2011a} (see also \cite{gitterman1995, choi2002}
for analogous series solutions, but focusing on the perturbation
regime). However, in that work only the first terms in the
expansion were explicitly given and higher order
terms were just outlined.\\
\indent In this work, based on the structure of the equations
obtained in \cite{urdapilleta2011a} for the survival probability,
we explicitly obtain all order functions. In particular, we find
the $n$th-term in a recursive scheme and prove by induction the
complete mathematical solution. From the survival probability, it
is straightforward to derive the FPT statistics, which maintains
the series structure. Since obtaining all order functions
(existence) does not imply the convergence of the series, in the
second part of the work, we analyse how does the expansion behave
in comparison with results obtained from simulations, as we
truncate the series at a finite order.

\section{Theoretical framework}
\indent In this section, we set the system under analysis, the
formalism we use to study the survival probability and the FPT
density function, and derive the complete solution.

\subsection{The system}
\indent The dynamics of the system is governed by the Langevin
equation

\begin{equation}\label{eq1}
   \frac{\rmd x}{\rmd t} = \mu +
   \frac{\epsilon}{\tau_{\rm d}}\rme^{-(t-t_{0})/\tau_{\rm d}}
   +\xi(t),
\end{equation}

\noindent where $x$ is the state variable (e.g. particle position,
membrane potential, etc), $t$ is the time, $\mu$ is the constant
(positive) component of the drift, $\epsilon$ and $\tau_{\rm d}$
characterize the intensity and the time constant of the
exponential time-dependent drift, respectively, $t_{0}$ sets the
initial time and $\xi(t)$ is a Gaussian white noise with
(constant) squared intensity $D$ [$\langle \xi(t) \rangle = 0$
and $\langle \xi(t) \xi(t') \rangle = 2D\delta(t-t')$].\\
\indent Following the derivation we have made in
\cite{urdapilleta2011a}, the probability that the particle remains
in the domain $x < x_{\rm thr}$ at time $t'$, given the (variable)
initial condition $x$ at (variable) time $t$, $F(t'|x,t)$, evolves
according to the backward FP equation

\begin{equation}\label{eq2}
   \frac{\partial F(t'|x,t)}{\partial t} = - \left[ \mu + \frac{\epsilon}{\tau_{\rm d}}~\rme^{-(t-t_{0})/\tau_{\rm d}}
   \right]\frac{\partial F(t'|x,t)}{\partial x} - D~\frac{\partial^2 F(t'|x,t)}{\partial
   x^2}.
\end{equation}

\indent In Eq.~(\ref{eq2}), $t'$ is a parameter accounting for the
present time. By making the substitution $\tau = t' - t$ and
renaming the probability as $F(x,\tau;t')$, Eq.~(\ref{eq2}) can be
written as

\begin{equation}\label{eq3}
   \frac{\partial F(x,\tau;t')}{\partial \tau} = \left[ \mu +
   \frac{\epsilon}{\tau_{\rm d}}~\rme^{-(t'-t_{0})/\tau_{\rm d}}~\rme^{\tau/\tau_{\rm d}}
   \right]\frac{\partial F(x,\tau;t')}{\partial x} + D~\frac{\partial^2 F(x,\tau;t')}{\partial x^2},
\end{equation}

\noindent which represents the equation to be solved. The system
is completed by specifying the initial and boundary conditions
\cite{urdapilleta2011a}, which are

\begin{eqnarray}
   \label{eq4}
   F(x,\tau=0;t') &=& \cases{1 ~~{\rm for}~x<x_{\rm thr},\\
   0~~{\rm for}~x\geq x_{\rm thr},}\\
   \label{eq5}
   F(x=x_{\rm thr},\tau;t') &=& 0.
\end{eqnarray}

\indent Equation~(\ref{eq4}) indicates that the survival at
initial time is certain for a particle located in the domain of
interest, whereas Eq.~(\ref{eq5}) establishes that the particle is
not allowed to be in $x \geq x_{\rm thr}$ and, therefore,
$x=x_{\rm thr}$ is an absorbent boundary.\\
\indent By proposing a solution as an expansion in $\epsilon$,

\begin{equation}\label{eq6}
\fl \hspace{0.5cm}
   F(x,\tau;t') = F_{0}(x,\tau;t') + \epsilon~F_{1}(x,\tau;t') +
   \epsilon^{2}~F_{2}(x,\tau;t') + \dots =
   \sum_{n=0}^{\infty}\epsilon^{n}~F_{n}(x,\tau;t'),
\end{equation}

\noindent Equation~(\ref{eq3}) reads

\begin{eqnarray}\label{eq7}
\fl \hspace{1.5cm}
   \left[ \frac{\partial F_{0}}{\partial \tau} - \mu \frac{\partial F_{0}}{\partial x} - D \frac{\partial^{2} F_{0}}{\partial x^{2}}
   \right] \nonumber\\
   + \sum_{n=1}^{\infty} \epsilon^{n}~\left[\frac{\partial F_{n}}{\partial \tau} - \mu \frac{\partial F_{n}}{\partial x} - \frac{1}{\tau_{\rm d}}~\rme^{-(t'-t_{0})/\tau_{\rm d}}~\rme^{\tau/\tau_{\rm d}}~\frac{\partial F_{n-1}}{\partial x} - D \frac{\partial^{2} F_{n}}{\partial
   x^{2}}\right] = 0.
\end{eqnarray}

\indent Since we expect that all functions $F_{n}$ do not depend
on $\epsilon$, Eq.~(\ref{eq6}), the expressions between brackets
should be identically $0$. Obviously, this hypothesis is true if
we are able to find $F_{n}(x,\tau;t')$. Under this condition, the
complete solution for $F(x,\tau;t')$ is given by the system of
equations

\begin{eqnarray}\label{eq8}
   \frac{\partial F_{0}}{\partial \tau} - \mu~\frac{\partial F_{0}}{\partial
   x} - D~\frac{\partial^{2} F_{0}}{\partial x^{2}} &=& 0, \\
   \label{eq9} \frac{\partial F_{n}}{\partial \tau} - \mu~\frac{\partial F_{n}}{\partial
   x} - D~\frac{\partial^{2} F_{n}}{\partial x^{2}}
   &=& \frac{1}{\tau_{\rm d}}~\rme^{-(t'-t_{0})/\tau_{\rm d}}~\rme^{\tau/\tau_{\rm d}}~\frac{\partial F_{n-1}}{\partial
   x}, ~~{\rm for}~~n \geq 1.
\end{eqnarray}

\indent Consistently with our previous assumption, given the
arbitrariness of $\epsilon$, the non-homogeneous initial
condition, $F(x,\tau=0;t') = 1$ for $x < x_{\rm thr}$, should be
exclusively imposed to the zeroth-order function
$F_{0}(x,\tau=0;t')$. In detail, initial conditions are

\begin{eqnarray}\label{eq10}
   F_{0}(x,\tau=0;t') &=& \cases{1 ~~{\rm if}~
   x<x_{\rm thr},\\
   0 ~~{\rm if}~x\geq x_{\rm thr},}\\
   \label{eq11}
   F_{n}(x,\tau=0;t') &=& 0 ~~{\rm for}~ n\geq 1.
\end{eqnarray}

\indent Completing the description, the boundary condition reads
$F_{n}(x=x_{\rm thr},\tau;t') = 0$, for all $n$.

\subsection{Survival probability from the backward state}
To obtain exactly the survival probability at time $t'$ from the
backward state we have to solve all terms involved in the
expansion given by Eq.~(\ref{eq6}). In particular, each term
satisfies a certain equation, Eq.~(\ref{eq8}) or (\ref{eq9}), with
appropriate conditions. Due to the different mathematical
structure, we focus on the zeroth-order term, $F_{0}(x,\tau;t')$,
separately from all other superior terms, $F_{n}(x,\tau;t')$ for
$n>0$.

\subsubsection{Zeroth-order term.}
\indent This term corresponds to the survival probability at time
$t'$ of a Brownian particle (initially) located in $x$ at time
$t$, when the system is driven exclusively by a constant positive
drift $\mu$ (in our system, this is obtained with $\epsilon = 0$).
According to the preceding derivation, $F_{0}(x,\tau;t')$
satisfies Eq.~(\ref{eq8}) with the conditions given by
Eq.~(\ref{eq10}) and $F_{0}(x_{\rm thr},\tau;t') = 0$. Since
$F_{0}(x,\tau;t') = 0$ for $x \geq x_{\rm thr}$, we focus
exclusively on $x < x_{\rm thr}$; in this case, by Laplace
transforming Eq.~(\ref{eq8}), we obtain

\begin{equation}\label{eq12}
   s~\tilde{F}_{0}^{L}(x) - \mu~\frac{\rmd\tilde{F}_{0}^{L}(x)}{\rmd x}-
   D~\frac{\rmd^{2}\tilde{F}_{0}^{L}(x)}{\rmd x^2} = 1,
\end{equation}

\noindent where $\tilde{F}_{0}^{L}(x,s;t') = \int_{0}^{\infty}
\rme^{-s\tau}~F_{0}(x,\tau;t')~d\tau$ is the Laplace transform of
$F_{0}(x,\tau;t')$ in the variable $\tau$. Since $t'$ and $s$ act
as parameters in Eq.~(\ref{eq12}), we have simplified the notation
to $\tilde{F}_{0}^{L}(x)$. In Laplace domain, the boundary
condition simply transforms to $\tilde{F}_{0}^{L}(x_{\rm thr}) = 0$.\\
\indent The solution to Eq.~(\ref{eq12}), with the preceding
condition and taking into account that $\tilde{F}_{0}^{L}(x)$
remains bounded as $x \rightarrow -\infty$, is

\begin{equation}\label{eq13}
   \tilde{F}_{0}^{L}(x) = \frac{1}{s} -
   \frac{1}{s}~\exp\left\{\frac{(x_{\rm thr}-x)}{2D}[\mu-\sqrt{\mu^{2}+4Ds}]\right\}.
\end{equation}

\indent The inverse Laplace transform of
$\tilde{F}_{0}^{L}(x,s;t')$ can be explicitly computed and reads

\begin{eqnarray}\label{eq14}
   F_{0}(x,\tau;t') = 1 &-& \frac{1}{2}~{\rm erfc} \Big[ \frac{(x_{\rm thr}-x)}{2\sqrt{D \tau}} - \frac{\mu}{2}\sqrt{\frac{\tau}{D}}\Big]\nonumber\\
   &-&\frac{1}{2}~\exp\Big[{\frac{(x_{\rm thr}-x)~\mu}{D}}\Big]~{\rm erfc} \Big[ \frac{(x_{\rm thr}-x)}{2\sqrt{D \tau}} + \frac{\mu}{2}\sqrt{\frac{\tau}{D}}\Big],
\end{eqnarray}

\noindent where ${\rm erfc}(x)$ is the complementary error
function.

\subsubsection{Higher order terms.}
\indent The equation governing the dynamics of the $n$th-order
function is given by

\begin{equation}\label{eq15}
\fl
   \frac{\partial F_{n}(x,\tau;t')}{\partial \tau} - \mu~\frac{\partial F_{n}(x,\tau;t')}{\partial
   x} - D~\frac{\partial^{2} F_{n}(x,\tau;t')}{\partial x^{2}} =
   \frac{1}{\tau_{\rm d}}
   \rme^{-(t'-t_{0})/\tau_{\rm d}}~\frac{\partial}{\partial
   x}\left[ \rme^{\tau/\tau_{\rm d}}~F_{n-1}(x,\tau;t')
   \right],
\end{equation}

\noindent and the boundary and initial conditions are
$F_{n}(x_{\rm thr},\tau;t')=0$ and $F_{n}(x,\tau=0;t') = 0$,
respectively.\\
\indent This equation can be solved via a Laplace transformation
in the variable $\tau$, which reads

\begin{equation}\label{eq16}
   s ~\tilde{F}_{n}^{L}(x) - \mu~\frac{\rmd \tilde{F}_{n}^{L}(x)}{\rmd x}
   - D~\frac{\rmd^{2}\tilde{F}_{n}^{L}(x)}{\rmd x^{2}} =
   \frac{1}{\tau_{\rm d}}~\rme^{-(t'-t_{0})/\tau_{\rm d}}~\frac{\rmd}{\rmd x}\left[
   \tilde{F}_{n-1}^{L}(x)\Big\rfloor_{s-1/\tau_{\rm
   d}}\right],
\end{equation}

\noindent where $\tilde{F}_{n}^{L}(x,s;t')$ is the Laplace
transform of $F_{n}(x,\tau;t')$, and its notation has been
simplified to $\tilde{F}_{n}^{L}(x)$ as in the previous case. Due
to the exponential pre-factor, the Laplace transform of the
forcing term has to be evaluated in the shifted variable
$s-1/\tau_{\rm d}$. Again, the boundary condition is simply
$\tilde{F}_{n}^{L}(x_{\rm thr}) = 0$.\\
\indent Next, we prove that the $n$th-order function is

\begin{eqnarray}\label{eq17}
\fl \hspace{1cm}
   \tilde{F}_{n}^{L}(x) = \rme^{-n(t'-t_{0})/\tau_{\rm d}}
   \frac{[\mu-\sqrt{\mu^{2}+4D(s-n/\tau_{\rm d})}]}{2D(s-n/\tau_{\rm d})}\nonumber\\
   \times \sum_{i=0}^{n} a_{n,i}(s)
   ~\exp\left\{\frac{(x_{\rm thr}-x)}{2D}[\mu-\sqrt{\mu^2+4D(s-i/\tau_{\rm d})}]\right\},
\end{eqnarray}

\noindent where the $n+1$ coefficients weighting each of the
exponential terms appearing in the solution of the $n$th-order
function, $a_{n,i}(s)$ ($i=0,\dots,n$), are given by

\begin{eqnarray}
   \label{eq18}
   \fl a_{n,0}(s) = \sum_{i=1}^{n}
   \frac{a_{n-1,i-1}(s-1/\tau_{\rm d})}{i}~\frac{[\mu-\sqrt{\mu^{2}+4D(s-i/\tau_{\rm d})}]}{2D},\\
   \label{eq19}
   \fl a_{n,i}(s) =
   -\frac{a_{n-1,i-1}(s-1/\tau_{\rm d})}{i}~\frac{[\mu-\sqrt{\mu^{2}+4D(s-i/\tau_{\rm d})}]}{2D},~~{\rm for}~i=1,\dots,n,
\end{eqnarray}

\noindent which build a recursive solution. In particular, from
Eq.~(\ref{eq19}) it is easy to check that $a_{n,0}(s) = -
\sum_{i=1}^{n} a_{n,i}(s)$.\\
\indent To demonstrate this solution, we will set
$\tilde{F}_{n-1}^{L}(x)$ according to the preceding proposition
and prove that the following order function satisfies the same
structure, Eq.~(\ref{eq17}). In doing so, we will derive
explicitly the recursive scheme given by Eqs.~(\ref{eq18}) and
(\ref{eq19}). The demonstration will be completed by showing that
the first-order term, $\tilde{F}_{1}^{L}(x)$, belongs to the
family of functions defined by Eq.~(\ref{eq17}) (i.e. we
prove the proposition by mathematical induction).\\
\indent Given that $\tilde{F}_{n-1}^{L}(x)$ is expressed according
to Eq.~(\ref{eq17}),

\begin{eqnarray}\label{eq20}
\fl \hspace{1cm}
   \tilde{F}_{n-1}^{L}(x) =
   \rme^{-(n-1)(t'-t_{0})/\tau_{\rm d}}~\frac{\{\mu-\sqrt{\mu^{2}+4D[s-(n-1)/\tau_{\rm d}]}\}}{2D[s-(n-1)\tau_{\rm d}]}
   \hspace{2.1cm}\nonumber\\
   \times \sum_{i=0}^{n-1} a_{n-1,i}(s)~\exp\left\{
   \frac{(x_{\rm thr}-x)}{2D}[\mu-\sqrt{\mu^{2}+4D(s-i/\tau_{\rm d})}]
   \right\},
\end{eqnarray}

\noindent the next order function, $\tilde{F}_{n}^{L}(x)$, is
given as the solution of Eq.~(\ref{eq16}) with an explicit forcing
term,

\begin{eqnarray}\label{eq21}
\fl \hspace{1cm}
   s ~\tilde{F}_{n}^{L}(x) - \mu~\frac{\rmd \tilde{F}_{n}^{L}(x)}{\rmd x}
   - D~\frac{\rmd^{2}\tilde{F}_{n}^{L}(x)}{\rmd x^{2}} =
   -\rme^{-n(t'-t_{0})/\tau_{\rmd}}~\frac{[\mu-\sqrt{\mu^{2}+4D(s-n/\tau_{\rmd})}]}
   {2D~\tau_{\rmd}~(s-n/\tau_{\rmd})}\nonumber\\
   \times \sum_{i=0}^{n-1} a_{n-1,i}(s-1/\tau_{\rmd})~\frac{\{\mu-\sqrt{\mu^{2}+4D[s-(i+1)/\tau_{\rmd}]}\}}{2D}\nonumber\\
   \times ~\exp\left\{\frac{(x_{\rm thr}-x)}{2D}~\left\{\mu-\sqrt{\mu^{2}+4D[s-(i+1)/\tau_{\rmd}]}\right\}\right\}.
\end{eqnarray}

\indent The solution to the homogeneous part of this equation
reads

\begin{equation}\label{eq22}
\fl \hspace{1cm}
   \tilde{F}_{n,{\rm hom}}^{L}(x) = C_{1}~\exp\left\{
   \frac{-\mu+\sqrt{\mu^{2}+4Ds}}{2D}~x \right\} + C_{2}~\exp\left\{
   \frac{-\mu-\sqrt{\mu^{2}+4Ds}}{2D}~x \right\},
\end{equation}

\noindent whereas it is easy to check that a particular solution
is

\begin{eqnarray}\label{eq23}
\fl \hspace{1cm}
   \tilde{F}_{n,{\rm part}}^{L}(x) = -
   \rme^{-n(t'-t_{0})/\tau_{\rm d}}~\frac{[\mu-\sqrt{\mu^{2}+4D(s-n/\tau_{\rm d})}]}{2D(s-n/\tau_{\rm d})}\nonumber\\
   \times \sum_{i=0}^{n-1} \frac{a_{n-1,i}(s-1/\tau_{\rm d})}{i+1}~\frac{\{ \mu - \sqrt{\mu^{2}+4D[s-(i+1)/\tau_{\rm d}]}\}}{2D}\nonumber\\
   \times ~\exp\left\{ \frac{(x_{\rm thr}-x)}{2D}~\left\{ \mu - \sqrt{\mu^{2}+4D[s-(i+1)/\tau_{\rm d}]}
   \right\}\right\}.
\end{eqnarray}

\indent The general solution is obtained from the combination of
Eqs.~(\ref{eq22}) and (\ref{eq23}), $\tilde{F}_{n}^{L}(x) =
\tilde{F}_{n,{\rm hom}}^{L}(x) + \tilde{F}_{n,{\rm part}}^{L}(x)$,
and it is valid for ${\rm Re}(s) \geq n/\tau_{\rm d}$. Since
$\tilde{F}_{n}^{L}(x)$ is bounded as $x \rightarrow -\infty$,
$C_{2}$ is $0$; at the same time, the boundary condition,
$\tilde{F}_{n}^{L}(x_{\rm thr}) = 0$, builds

\begin{eqnarray}\label{eq24}
\fl \hspace{1cm}
   C_{1} = \rme^{-n(t'-t_{0})/\tau_{\rm d}}~\frac{[\mu-\sqrt{\mu^{2}+4D(s-n/\tau_{\rm d})}]}{2D(s-n/\tau_{\rm d})}~\exp
   \left( \frac{\mu-\sqrt{\mu^{2}+4Ds}}{2D} ~ x_{\rm thr}\right)\nonumber\\
   \times \sum_{i=0}^{n-1}\frac{a_{n-1,i}(s-1/\tau_{\rm d})}{i+1}~\frac{\{ \mu - \sqrt{\mu^{2}+4D[s-(i+1)/\tau_{\rm d}]}
   \}}{2D}.
\end{eqnarray}

\indent Therefore, the $n$th-order function is given by

\begin{eqnarray}\label{eq25}
\fl \hspace{1cm}
   \tilde{F}_{n}^{L}(x) = \rme^{-n(t'-t_{0})/\tau_{\rm d}}~\frac{[\mu-\sqrt{\mu^{2}+4D(s-n/\tau_{\rm d})}]}{2D(s-n/\tau_{\rm d})}\nonumber\\
   \times \Bigg\{a_{n,0}(s)~\exp\left\{\frac{(x_{\rm thr}-x)}{2D}~[\mu-\sqrt{\mu^{2}+4Ds}]\right\}\nonumber\\
   + \sum_{i=0}^{n-1} a_{n,i+1}(s)~\exp\left\{ \frac{(x_{\rm thr}-x)}{2D}~\left\{\mu-\sqrt{\mu^{2}+4D[s-(i+1)/\tau_{\rm d}]}\right\} \right\} \Bigg\},
\end{eqnarray}

\noindent where the coefficients appearing in Eq.~(\ref{eq25}) are

\begin{eqnarray}\label{eq26}
\fl
   a_{n,0}(s) = \sum_{i=0}^{n-1}
   \frac{a_{n-1,i}(s-1/\tau_{\rm d})}{i+1}~\frac{\{ \mu -
   \sqrt{\mu^{2}+4D[s-(i+1)/\tau_{\rm d}]}\}}{2D}\\
\fl
   \label{eq27}
   a_{n,i+1}(s) = -\frac{a_{n-1,i}(s-1/\tau_{\rm d})}{i+1}~\frac{\{ \mu - \sqrt{\mu^{2}+4D[s-(i+1)/\tau_{\rm
   d}]}\}}{2D},~~i=0,\dots,n-1.
\end{eqnarray}

\indent By shifting the index $i$ in the sum symbol, it is easy to
check that Eq.~(\ref{eq25}) is equal to Eq.~(\ref{eq17}), and each
of the coefficients, Eq.~(\ref{eq26}) or Eq.~(\ref{eq27}), is
given by Eq.~(\ref{eq18}) or Eq.~(\ref{eq19}), respectively.\\
\indent The proof is completed by showing that the first-order
function, given as the solution to Eq.~(\ref{eq16}) for $n=1$ and
$\tilde{F}_{1}^{L}(x_{\rm thr})=0$, is part of the family of
functions described by Eq.~(\ref{eq17}). As shown in
\cite{urdapilleta2011a}, this solution reads

\begin{eqnarray}\label{eq28}
\fl
   \tilde{F}_{1}^{L}(x) =
   \rme^{-(t'-t_{0})/\tau_{\rm d}}~\frac{[\mu-\sqrt{\mu^{2}+4D(s-1/\tau_{\rm d})}]}{2D(s-1/\tau_{\rm d})}
   ~\Bigg\{ -\exp\left\{ \frac{(x_{\rm thr}-x)}{2D}[\mu-\sqrt{\mu^{2}+4Ds}]\right\} \nonumber\\
   + ~\exp\left\{\frac{(x_{\rm thr}-x)}{2D}[\mu-\sqrt{\mu^{2}+4D(s-1/\tau_{\rm d})}]\right\}\Bigg\},
\end{eqnarray}

\noindent which can be easily checked satisfying Eq.~(\ref{eq17}).
Furthermore, from this solution we can observe that the
coefficients $a_{n,i}(s)$ ($n=2,\dots,\infty$ and $i=0,\dots,n$)
are recursively built from $a_{1,0}(s) = -1$ and
$a_{1,1}(s) = 1$.\\
\indent Even when not explicitly available, the $n$th-order
function in the temporal domain, $F_{n}(x,\tau;t')$, is given by
the inverse Laplace transform of Eq.~(\ref{eq17}), which reads

\begin{eqnarray}\label{eq29}
\fl \hspace{1cm}
   F_{n}(x,\tau;t') =
   \rme^{-n(t'-t_{0})/\tau_{\rm d}}~\frac{1}{2\pi {\rm j}}
   \int_{\sigma-{\rm j}\infty}^{\sigma+{\rm j}\infty}\rme^{s\tau}\frac{[\mu-\sqrt{\mu^{2}+4D(s-n/\tau_{\rm d})}]}{2D(s-n/\tau_{\rm d})}\nonumber\\
   \times~\sum_{i=0}^{n} a_{n,i}(s)~\exp\left\{
   \frac{(x_{\rm thr}-x)}{2D}[\mu-\sqrt{\mu^{2}+4D(s-i/\tau_{\rm d})}]\right\} \rmd s,
\end{eqnarray}

\noindent where ${\rm j}$ represents the imaginary unit and the
region of convergence of the integrand requires that $\sigma \geq
n/\tau_{\rm d}$. From the substitutions $z = s - n/\tau_{\rm d}$
for the integration variable and $k=n-i$ for the index of the sum,
we obtain

\begin{eqnarray}\label{eq30}
\fl \hspace{1cm}
   F_{n}(x,\tau;t') =
   \rme^{-n(t'-t_{0})/\tau_{\rm d}}~\rme^{n\tau/\tau_{\rm d}}~\frac{1}{2\pi
   {\rm j}}\int_{\sigma_{z}-{\rm j}\infty}^{\sigma_{z}+{\rm j}\infty}
   \rme^{z\tau}~\frac{[\mu-\sqrt{\mu^{2}+4Dz}]}{2Dz}\nonumber\\
   \times~\sum_{k=0}^{n}b_{n,k}(z)~\exp\left\{
   \frac{(x_{\rm thr}-x)}{2D}[\mu-\sqrt{\mu^{2}+4D(z+k/\tau_{\rm d})}]\right\} \rmd z,
\end{eqnarray}

\noindent where now, $\sigma_{z}\geq 0$. In Eq.~(\ref{eq30}), we
have defined new coefficients for the exponential terms appearing
in the sum symbol, $b_{n,k}(z) = a_{n,n-k}(z+n/\tau_{\rm d})$.
With this definition, the recursive structure is

\begin{eqnarray}\label{eq31}
\fl
   b_{n,k}(z) &=&
   -\frac{b_{n-1,k}(z)}{n-k}~\frac{[\mu-\sqrt{\mu^{2}+4D(z+k/\tau_{\rm d})}]}{2D},~~{\rm for}~k=0,\dots,n-1,\\
\fl
   \label{eq32}
   b_{n,n}(z) &=& -\sum_{k=0}^{n-1} b_{n,k}(z),
\end{eqnarray}

\noindent starting from $b_{1,0}(z)=1$ and $b_{1,1}(z)=-1$.

\subsection{Survival probability}
\indent The survival probability of the particle at time $t'$
arises when we impose the initial state to the backward state, $x
= x_{0}$ at time $t = t_{0}$. As in the previous subsection, we
discriminate between the zeroth-order term from all superior order
functions.

\subsubsection{Zeroth-order term.}
This term is given by imposing the initial state in
Eq.~(\ref{eq14}) and reads

\begin{eqnarray}\label{eq33}
   F_{0}(\tau) = 1 &-& \frac{1}{2}~{\rm erfc} \Big[ \frac{(x_{\rm thr}-x_{0})}{2\sqrt{D \tau}} - \frac{\mu}{2}\sqrt{\frac{\tau}{D}}\Big]\nonumber\\
   &-&\frac{1}{2}~\exp\Big[{\frac{(x_{\rm thr}-x_{0})~\mu}{D}}\Big]~{\rm erfc} \Big[ \frac{(x_{\rm thr}-x_{0})}{2\sqrt{D \tau}} + \frac{\mu}{2}\sqrt{\frac{\tau}{D}}\Big],
\end{eqnarray}

\noindent where now, $\tau = t' - t_{0}$ is the actual time
difference (time elapsed from the initial time $t_0$ to the
present time $t'$). Note that we have eliminated the dependence on
$t'$ in the notation for $F_{0}(\tau)$, since it only appears in
the combination given by $\tau$.\\
\indent The Laplace transform of $F_{0}(\tau)$ in the variable
$\tau$ is given by

\begin{equation}\label{eq34}
   \tilde{F}_{0}^{L}(s) = \frac{1}{s} -
   \frac{1}{s}~\exp\left\{\frac{(x_{\rm
   thr}-x_{0})}{2D}[\mu-\sqrt{\mu^{2}+4Ds}]\right\},
\end{equation}

\noindent which is one of the quantities of interest for the
assessment of the FPT density function.

\subsubsection{Higher order terms.}
\indent As in the zeroth-order, these terms are obtained from the
evaluation of the initial state in the corresponding expression
for the survival probability from the backward state,
Eq.~(\ref{eq30}), which yields

\begin{eqnarray}\label{eq35}
\fl \hspace{1cm}
   F_{n}(\tau) = \frac{1}{2\pi {\rm j}} \int_{\sigma_{z}-{\rm j}\infty}^{\sigma_{z}+
   {\rm j}\infty} \rme^{z\tau}~\frac{[\mu-\sqrt{\mu^{2}+4Dz}]}{2Dz}\nonumber\\
   \times~\sum_{k=0}^{n}b_{n,k}(z)~\exp\left\{
   \frac{(x_{\rm thr}-x_{0})}{2D}[\mu-\sqrt{\mu^{2}+4D(z+k/\tau_{\rm d})}]\right\} \rmd z,
\end{eqnarray}

\noindent where $\tau = t'- t_{0}$ is the actual time difference
and the dependence on $t'$ appears only through $\tau$.\\
\indent The Laplace transform of $F_{n}(\tau)$ in the variable
$\tau$ is readily obtained, and reads

\begin{eqnarray}\label{eq36}
\fl \hspace{1cm}
   \tilde{F}^{L}_{n}(s) = \frac{[\mu-\sqrt{\mu^{2}+4Ds}]}{2Ds}\nonumber\\
   \times~\sum_{k=0}^{n}b_{n,k}(s)~\exp\left\{
   \frac{(x_{\rm thr}-x_{0})}{2D}[\mu-\sqrt{\mu^{2}+4D(s+k/\tau_{\rm d})}]\right\},
\end{eqnarray}

\noindent where the coefficients $b_{n,k}(s)$ are given by
Eqs.~(\ref{eq31}) and (\ref{eq32}) in the variable $s$.

\subsection{First-passage-time statistics}
The probability that a Brownian particle driven by an exponential
time-dependent drift (superimposed to a linear field) remains in
the domain $x<x_{\rm thr}$ at time $t'$, having started at time
$t_{0}$ in the position $x_{0}$, is given by the survival
probability calculated in the previous subsection, $F(\tau)$. It
was demonstrated that this probability can be written as a series

\begin{equation}\label{eq37}
   F(\tau) = \sum_{n=0}^{\infty} \epsilon^{n} ~ F_{n}(\tau),
\end{equation}

\noindent where each term depends exclusively on the variable
$\tau = t'-t_{0}$. The different order terms are obtained from the
functions explicitly found in Eqs.~(\ref{eq34}) and
(\ref{eq36}), in the Laplace domain.\\
\indent For a positive $\mu$, the particle will cross the level
$x=x_{\rm thr}$ for the first time at time $T$ (and will be
absorbed), and this random variable represents the FPT. Given the
survival probability at time $\tau$ (time elapsed from time
$t_{0}$), the FPT for this particle satisfies $T > \tau$ (it is
absorbed at a posterior time); therefore, the survival probability
represents $F(\tau) = {\rm Prob}(T > \tau)$ and the cumulative
distribution function for the FPT, $\Phi(\tau)$, is given by
$\Phi(\tau)=1-F(\tau)$. Consequently, the density function for the
FPT, $\phi(\tau)$, is

\begin{equation}\label{eq38}
   \phi(\tau) = \frac{\rmd \Phi(\tau)}{\rmd \tau} = - \frac{\rmd F(\tau)}{\rmd \tau}.
\end{equation}

\indent Since $F(\tau)$ is given as a series solution,
Eq.~(\ref{eq37}), the density function $\phi(\tau)$ can also be
expressed as a series,

\begin{equation}\label{eq39}
   \phi(\tau) = \sum_{n=0}^{\infty} \epsilon^{n}~\phi_{n}(\tau),
\end{equation}

\noindent where the functions $\phi_{n}(\tau)$ are

\begin{equation}\label{eq40}
   \phi_{n} = -\frac{\rmd F_{n}(\tau)}{\rmd \tau}.
\end{equation}

\indent In Laplace domain, the preceding series is expressed as

\begin{equation}\label{eq41}
   \tilde{\phi}^{L}(s) = \sum_{n=0}^{\infty} \epsilon^{n}~\tilde{\phi}^{L}_{n}(s),
\end{equation}

\noindent where \cite{urdapilleta2011a}

\begin{eqnarray}
   \label{eq42}
   \tilde{\phi}_{0}^{L}(s) = 1 - s~\tilde{F}_{0}^{L}(s),\\
   \label{eq43}
   \tilde{\phi}_{n}^{L}(s) = - s~\tilde{F}_{n}^{L}(s), ~~{\rm
   for}~n\geq 1.
\end{eqnarray}

\indent Replacing the results we have obtained in the previous
subsection, Eqs.~(\ref{eq34}) and (\ref{eq36}), these functions
explicitly read

\begin{eqnarray}
\label{eq44} \fl \hspace{1cm}
   \tilde{\phi}_{0}^{L}(s) = \exp\left\{\frac{(x_{\rm
   thr}-x_{0})}{2D}[\mu-\sqrt{\mu^{2}+4Ds}]\right\},\\
\label{eq45} \fl \hspace{1cm}
   \tilde{\phi}_{n}^{L}(s) = - \frac{[\mu-\sqrt{\mu^{2}+4Ds}]}{2D}\nonumber\\
\fl \hspace{2.2cm}
   \times~\sum_{k=0}^{n}b_{n,k}(s)~\exp\left\{
   \frac{(x_{\rm thr}-x_{0})}{2D}[\mu-\sqrt{\mu^{2}+4D(s+k/\tau_{\rm
   d})}]\right\},~~{\rm for}~n\geq 1,
\end{eqnarray}

\noindent where the coefficients $b_{n,k}(s)$ are given by
Eqs.~(\ref{eq31}) and (\ref{eq32}) in the variable $s$, and the
recursive structure starts from $b_{1,0}(s)=1$ and
$b_{1,1}(s)=-1$. In order to exemplify this recursive
construction, in table \ref{tabl} we show the coefficients
associated with the exponential terms, $b_{n,k}(s)$, up to the
fourth order.

\section{Comparison to numerical simulations}
\begin{figure}[t]
\begin{center}
\includegraphics[scale=0.5]{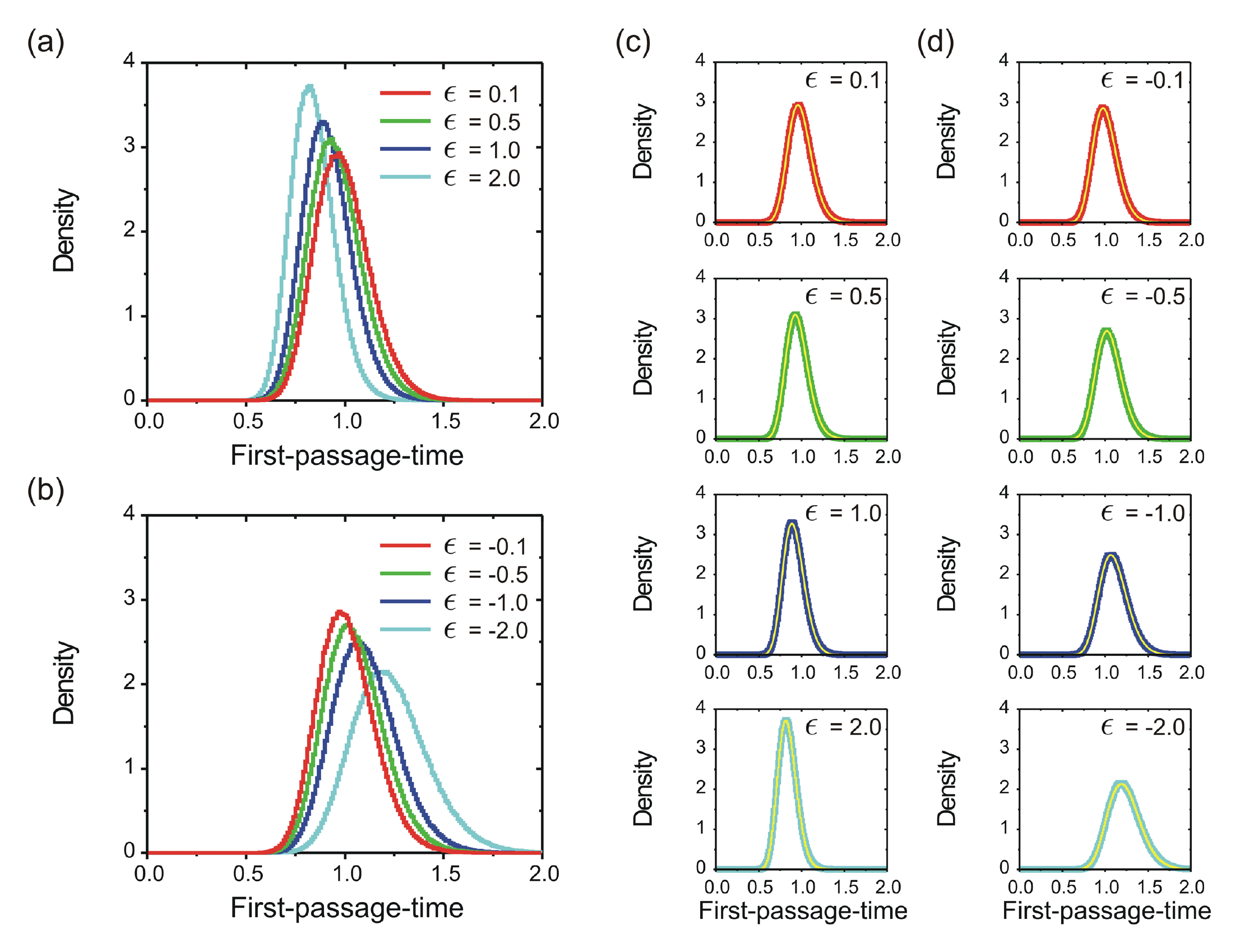}
\caption{\label{fig1} Comparison between the series solution for
the FPT density function and numerical results. (a) and (b)
First-passage-time distributions obtained from the numerical
simulation of Eq.~(\ref{eq1}), for different (a) positive and (b)
negative intensities of the time-dependent drift $\epsilon$
(colored histograms). (c) Each histogram shown in (a) is properly
described by the series solution, Eq.~(\ref{eq37}), shown as a
thin yellow line. (d) Equivalent comparison between theoretical
results and the histograms shown in (b), for negative intensities.
In each case, the series is truncated in an adequate order, $N$:
$\phi(\tau) = \sum_{n=0}^{N} \epsilon^{n}~\phi_{n}(\tau)$. As the
value of $\epsilon$ increases in magnitude, the order used to
represent the theoretical result increases as well. In particular,
$N=1$ for $\epsilon = \pm 0.1$ (top), $N=2$ for $\epsilon = \pm
0.5$ (middle-top), $N=4$ for $\epsilon = \pm 1.0$ (middle-bottom),
and $N=9$ for $\epsilon = \pm 2.0$ (bottom). Parameters: $\mu =
1$, $x_{\rm thr}-x_{0} = 1$, $D = 0.01$, and $\tau_{\rm d}= 10$.}
\end{center}
\end{figure}

\begin{figure}[t]
\begin{center}
\includegraphics[scale=0.7]{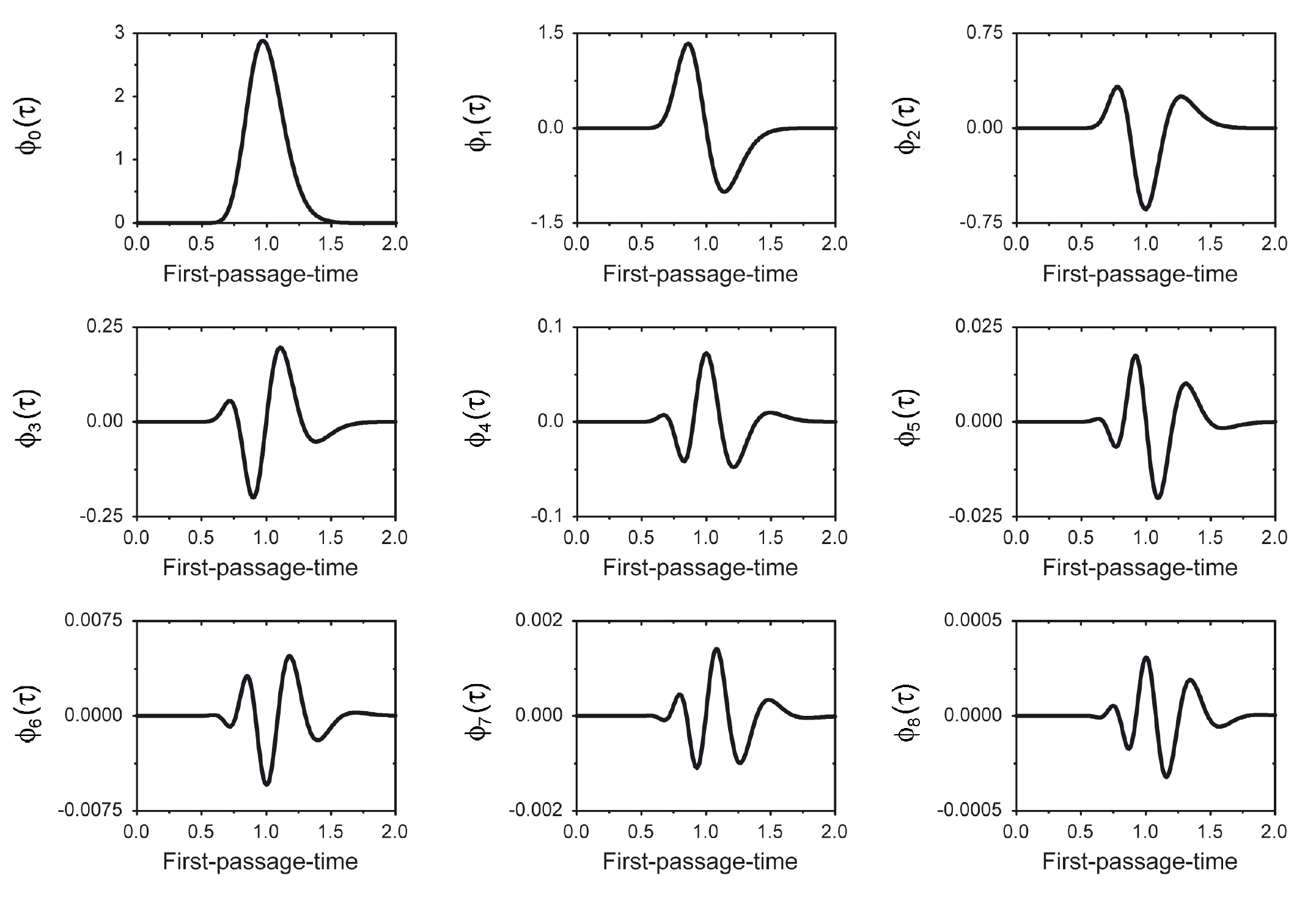}
\caption{\label{fig2} Functions $\phi_{n}(\tau)$ used to construct
the theoretical description of the FPT statistics for the test
case defined in Fig.~(\ref{fig1}). Note that the $y$-scale varies
from panel to panel and, particularly, decreases as the order
becomes higher.}
\end{center}
\end{figure}

\indent To illustrate the solution we have obtained for the FPT
problem of a Brownian particle driven by an exponential
time-dependent drift, in this section we compare the analytical
results with data extracted from numerical simulations. Samples of
the FPT distribution are collected from the times at which a
Brownian particle, evolving according to the Langevin equation
given by Eq.~(\ref{eq1}) and starting from $x_{0}$, arrives at the
threshold $x_{\rm thr}$ for the first time. As we have shown in
\cite{urdapilleta2011a}, for small intensities of the
time-dependent drift, $\epsilon$, the system corresponds to a
perturbation scenario, and the first-order solution, $\phi(\tau) =
\phi_{0}(\tau) + \epsilon~\phi_{1}(\tau)$, properly describes the
FPT statistics. In this section, we extend that comparison beyond
the linear regime, for large values of $\epsilon$. Obviously, in
this case we need to include higher order terms in the series
given by Eq.~(\ref{eq37}). This comparison is not merely
illustrative; since the characterization of the series convergence
remains elusive to us, we resort to a numerical test case to
demonstrate the usefulness of the series solution.\\
\indent Considering $\epsilon = 0$, the dynamics defined by
Eq.~(\ref{eq1}) has no intrinsic timescale and, therefore, time
units can be normalized by $(x_{\rm thr}-x_{0})/\mu$ (i.e. the
external parameter $\mu$ defines the escape rate). Additionally,
we consider the non-dimensional form of this equation, obtained by
setting $x/(x_{\rm thr}-x_{0}) \rightarrow x$, $\epsilon/(x_{\rm
thr}-x_{0}) \rightarrow \epsilon$, and $D/[(x_{\rm thr}-x_{0})\mu]
\rightarrow D$. The preceding procedures are equivalent to set
$\mu = 1$ and $x_{\rm thr}-x_{0} = 1$ in the system described by
Eq.~(\ref{eq1}). Given our interest in neural adaptation, the
remaining parameters will be defined from typical values in
adapting neurons. For the system without the time-dependent drift,
$\epsilon = 0$, the mean FPT is $\langle \tau \rangle = 1$; in the
context we focus on, a proper scale for $\tau_{\rm d}$ is about
$10$ times this value \cite{benda2003}, $\tau_{\rm d} = 10$. Since
the squared noise intensity strongly influences the dispersion of
the interspike interval distribution (FPT statistics), its value
is selected to produce typical histograms obtained in experiments
\cite{gerstein1964}, $D = 0.01$. The intensity of the
time-dependent drift weights the influence of an adaptation
current in the intrinsic FPT distribution (in particular, in the
firing rate) and constitutes a negative feedback to the
subthreshold integration \cite{urdapilleta2011b} ($\epsilon < 0$);
given that our interest here is to analyse the behavior of the
series solution, this parameter will be used to set different
regimes beyond the linear case.\\
\indent In Figs.~(\ref{fig1}.a) and (\ref{fig1}.b) we show the FPT
density function constructed from numerical data, for different
intensities of the time-dependent exponential drift, $\epsilon$.
As expected, for positive (negative) intensities, as the strength
of the time-dependent drift increases in magnitude, the threshold
is reached at earlier (later) times and, consequently, the FPT
distribution shifts towards lower (larger) values. In
Figs.~(\ref{fig1}.c) and (\ref{fig1}.d), the different
distributions are separately compared with analytical results.
These results are based on the truncated series solution [see
Eq.~(\ref{eq37})], $\phi(\tau) = \sum_{n=0}^{N}
\epsilon^{n}~\phi_{n}(\tau)$, where $N$ is selected to reproduce
numerical data. As shown in Figs.~(\ref{fig1}.c) and
(\ref{fig1}.d) (top panel), the FPT distribution for $\epsilon =
\pm 0.1$ is precisely described by the linear expansion
(perturbation regime), $\phi(\tau) = \phi_{0}(\tau) \pm
\epsilon~\phi_{1}(\tau)$. However, the proper description of the
numerical distributions for higher intensities requires the
addition of higher order terms. For example, as shown in
Figs.~(\ref{fig1}.c) and (\ref{fig1}.d), FPT distributions for
$\epsilon = \pm 0.5$, $\epsilon = \pm 1.0$, and $\epsilon = \pm
2.0$ are described with $N = 2$, $N = 4$, and $N = 9$,
respectively (middle-top, middle-bottom, and bottom panels,
respectively).\\
\indent Except for the zeroth-order term, which can be explicitly
computed in the temporal domain via the inverse Laplace
transformation of Eq.~(\ref{eq44}) and known as the inverse
Gaussian distribution \cite{gerstein1964},

\begin{equation}\label{eq46}
   \phi_{0}(\tau) = \frac{x_{\rm thr}-x_{0}}{\sqrt{4 \pi D\tau^{3}}}~\exp\left\{ -\frac{\left[ (x_{\rm
thr} - x_{0}) - \mu\tau \right]^{2}}{4D\tau}\right\},
\end{equation}

\noindent all superior order functions, $\phi_{n}(\tau)$ for
$n\geq 1$, require the numerical (inverse Laplace) transform of
Eq.~(\ref{eq45}). In Fig.~(\ref{fig2}) we show these functions up
to the eighth order, for the parameters defined in
Fig.~(\ref{fig1}). It is worthwhile to note that these functions
decrease in amplitude as the order increases (see $y$-scales),
which is indicative of the convergence of the series (but not
conclusive). An additional point to take into account in this
analysis is that the numerical transform introduces an error which
limits the reliability of the results. In particular, for the test
case used here, the numerical inversion of the functions beyond
the tenth order is inaccurate and, therefore, the comparison
between analytical and numerical results is restricted to
$\epsilon \sim \pm 2.0$ [Figs.~(\ref{fig1}.c) and (\ref{fig1}.d)].\\
\indent This numerical inaccuracy can be circumvented if we
analyse properties that can be obtained directly from the Laplace
transform of the FPT density function, $\tilde{\phi}^{L}(s)$; for
example, its moments read \cite{urdapilleta2011a}

\begin{equation}\label{eq47}
   \langle \tau^{k} \rangle = \int_{0}^{\infty}
   \phi(\tau)~\tau^{k}~{\rm d}\tau = (-1)^{k}~\frac{{\rm d}^{k} \tilde{\phi}^{L}(s)}{{\rm
   d}s^{k}}\Big\rfloor_{s=0}.
\end{equation}

\indent It is easy to check that, due to the linear nature,
Eq.~(\ref{eq47}) adopts a series structure when
$\tilde{\phi}^{L}(s)$ is replaced by Eq.~(\ref{eq41}). Explicitly,
by defining

\begin{equation}\label{eq48}
   \langle \tau^{k} \rangle_{\phi_{n}} = (-1)^{k}~\frac{{\rm d}^{k} \tilde{\phi}_{n}^{L}(s)}{{\rm
   d}s^{k}}\Big\rfloor_{s=0},
\end{equation}

\noindent Equation~(\ref{eq47}) results in

\begin{equation}\label{eq48}
   \langle \tau^{k} \rangle = \sum_{n=0}^{\infty} \epsilon^{n}~\langle \tau^{k} \rangle_{\phi_{n}}.
\end{equation}

\begin{figure}[t]
\begin{center}
\includegraphics[scale=0.5]{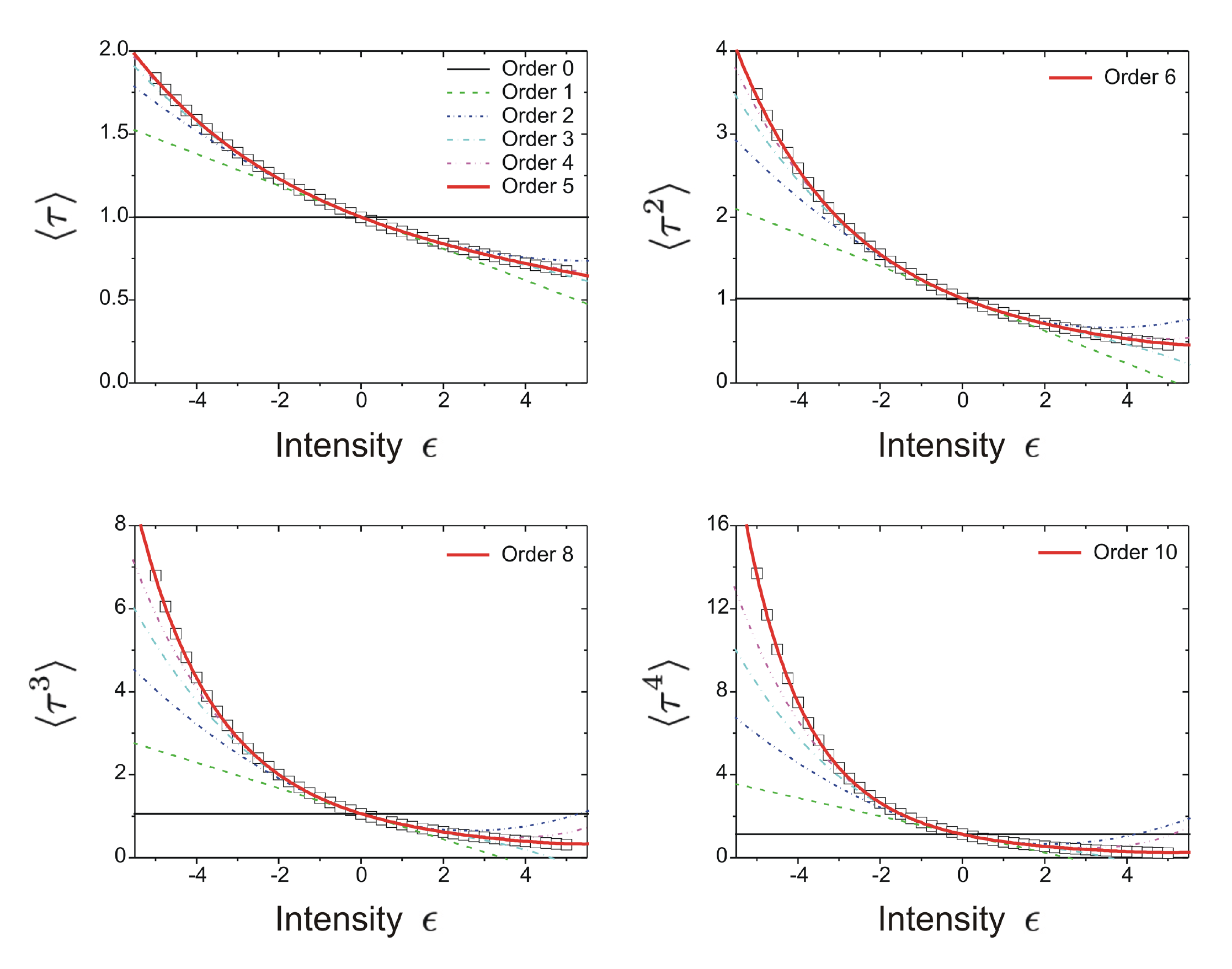}
\caption{\label{fig3} Comparison between numerical and theoretical
results for the moments of the FPT distribution, as a function of
the intensity of the time-dependent drift $\epsilon$. In all
cases, symbols represent averages of numerical results, whereas
lines correspond to Eq.~(\ref{eq48}) truncated at the order
indicated. Parameters are defined as in Fig.~(\ref{fig1}).}
\end{center}
\end{figure}

\indent In Fig.~(\ref{fig3}) we show a comparison between
numerical and analytical results for the first four moments as a
function of the intensity of the time-dependent drift. Parameters
are defined as those corresponding to the test case analysed
previously. This means that all $\langle \tau^{k}
\rangle_{\phi_{n}}$ are certain scalars and Eq.~(\ref{eq48})
represents a polynomial in $\epsilon$. It is interesting to note
that the order of the truncated polynomial, $N$, necessary to
describe the numerical results increases as the exponent of the
moment does. Alternatively, for a given order, the analytical
expressions for the lowest moments remain valid in a larger range
of $|\epsilon|$. Also, it is interesting to point out that
Eq.~(\ref{eq48}) implies a different behavior for positive or
negative values of $\epsilon$. As shown in Fig.~(\ref{fig3}), for
$\epsilon < 0$ the convergence of the analytical expression is
smooth as the order of the polynomial increases, whereas for
$\epsilon > 0$ the convergence exhibits an alternating
character.\\
\indent From the moments of the FPT density function we can obtain
other properties; in particular, its cumulants. In this case, the
expressions relating both properties should be developed, and a
series is obtained by grouping together equal order terms. For
example, the second cumulant is, up to the first order in
$\epsilon$, $\langle (\tau - \langle \tau \rangle)^{2} \rangle =
\left[ \langle \tau^{2} \rangle_{\phi_{0}} - \langle \tau
\rangle^{2}_{\phi_{0}} \right] + \epsilon \left[ \langle \tau^{2}
\rangle_{\phi_{1}} - 2 \langle \tau \rangle_{\phi_{0}} \langle
\tau \rangle_{\phi_{1}}\right] + \mathcal{O}(\epsilon^{2})$.\\

\section{Concluding remarks and discussion}
\indent We have studied the FPT statistics resulting from the
biased diffusion of a Brownian particle up to a threshold $x_{\rm
thr}$, when the constant drift is supplemented with an exponential
time-dependent component [see Eq.~(\ref{eq1})]. In a previous work
\cite{urdapilleta2011a}, we analysed this time-inhomogeneous
system in the backward FP formalism, derived the diffusion
equation governing the evolution of the survival probability from
the backward state, Eq.~(\ref{eq3}), and proposed a solution as a
series in terms of the intensity of the time-dependent drift,
Eq.~(\ref{eq6}). In that work we focused on a perturbation regime
and explicitly solved the expansion up to the first-order terms.
In this work, we have extended these results by explicitly
computing all superior order functions in a recursive scheme [see
Eq. (\ref{eq30})]. The survival probability (with the initial
state imposed) and the FPT statistics are easily derived from this
solution and preserve the series structure; for completeness,
their superior order terms are also explicitly given [see
Eqs.~(\ref{eq36}) and (\ref{eq45}), for the corresponding
expressions in the Laplace domain]. In the second part of this
work we have defined a test case in order to assess the usefulness
of the series solution. Analytical and numerical results are
compared for different intensities of the time-dependent drift
(beyond the perturbation regime), and a remarkable agreement is
found for each case whenever the series is
truncated in an adequate order.\\
\indent The problem we have analysed provides the intrinsic
statistics of the events defined by an adapting neuron (interspike
intervals). In this case, the system state corresponds to the
membrane potential and the exponential time-dependent drift
resembles a specific ionic current that decays during the
subthreshold integration. This kind of currents supports a widely
observed phenomenon in neurons, known as \textit{spike-frequency
adaptation} (SFA), when the initial state of the current (in the
present framework, proportional to $\epsilon$) is properly coupled
with the spiking history \cite{urdapilleta2011b}. Particularly,
they are restricted to be negative ($\epsilon < 0$), providing a
feedback to the neuron that lengths the interspike interval (FPT).
In this work, we have focused on the statistics describing a
single interspike interval for a given initial current [i.e. the
FPT statistics analysed here corresponds to a conditional
distribution, $\phi(\tau|\epsilon)$, in the history-dependent
spike train]. As shown in \cite{urdapilleta2011b}, the analysis of
the successive events in a neuron exhibiting SFA can be performed
with a hidden Markov model. In this case, the conditional
distribution is essential to study the spike train properties and
its explicit assessment has motivated the contribution made in
this study.

\section{Acknowledgments}
This work was supported by the Consejo de Investigaciones
Cient\'ificas y T\'ecnicas de la Rep\'ublica Argentina.

\newpage
\Table{\label{tabl}Coefficients $b_{n,k}(s)$ weighting the
exponential terms that compose the solution to the survival
probability and the FPT density function up to the fourth order.
For the sake of clarity, we have defined the auxiliary
coefficients $c_{n}(s) = [\mu-\sqrt{\mu^{2}+4D(s+n/\tau_{\rm
d})}]/(2D)$.} \br
Order&Coefficients\\
\br
$n=1$ & $b_{1,0}(s) = 1$\\
      & $b_{1,1}(s) = -1$\\
\mr
$n=2$ & $b_{2,0}(s) =  -[b_{1,0}(s)/2] c_{0}(s) = -\frac{1}{2} c_{0}(s)$\\
      & $b_{2,1}(s) = - [b_{1,1}(s)/1] c_{1}(s) = c_{1}(s)$\\
      & $b_{2,2}(s) = - b_{2,0}(s) - b_{2,1}(s) = \frac{1}{2} c_{0}(s) - c_{1}(s)$\\
\mr
$n=3$ & $b_{3,0}(s) = -[b_{2,0}(s)/3] c_{0}(s) = \frac{1}{6} [c_{0}(s)]^{2}$\\
      & $b_{3,1}(s) = -[b_{2,1}(s)/2] c_{1}(s) = -\frac{1}{2} [c_{1}(s)]^{2}$\\
      & $b_{3,2}(s) = -[b_{2,2}(s)/1] c_{2}(s) = [-\frac{1}{2}c_{0}(s)+c_{1}(s)]c_{2}(s)$\\
      & $b_{3,3}(s) = - b_{3,0}(s) - b_{3,1}(s) - b_{3,2}(s) = -\frac{1}{6} [c_{0}(s)]^{2} + \frac{1}{2}
      [c_{1}(s)]^{2} + [\frac{1}{2}c_{0}(s)-c_{1}(s)]c_{2}(s)$\\
\mr

$n=4$ & $b_{4,0}(s) = -[b_{3,0}(s)/4] c_{0}(s) = -\frac{1}{24} [c_{0}(s)]^{3}$\\
      & $b_{4,1}(s) = -[b_{3,1}(s)/3] c_{1}(s) = \frac{1}{6} [c_{1}(s)]^{3}$\\
      & $b_{4,2}(s) = -[b_{3,2}(s)/2] c_{2}(s) = \frac{1}{2}[\frac{1}{2} c_{0}(s) - c_{1}(s) ] [c_{2}(s)]^{2}$\\
      & $b_{4,3}(s) = -[b_{3,3}(s)/1] c_{3}(s) =
      \{\frac{1}{6}[c_{0}(s)]^{2}-\frac{1}{2}[c_{1}(s)]^{2}+[-\frac{1}{2}c_{0}(s)+c_{1}(s)]c_{2}(s)\}c_{3}(s)$\\
      & $b_{4,4}(s) = - b_{4,0}(s) - b_{4,1}(s) - b_{4,2}(s) - b_{4,3}(s)$\\
      & $\hspace{1.065cm} = \frac{1}{24} [c_{0}(s)]^{3} - \frac{1}{6}
      [c_{1}(s)]^{3} + \frac{1}{2}[-\frac{1}{2} c_{0}(s) +
      c_{1}(s)][c_{2}(s)]^{2}$\\
      & $\hspace{1.5cm} + \{-\frac{1}{6}[c_{0}(s)]^{2}+\frac{1}{2}[c_{1}(s)]^{2}+[\frac{1}{2}c_{0}(s)-c_{1}(s)]c_{2}(s)\}c_{3}(s)$\\

\br
\endTable

\newpage

\newpage

\newpage

\end{document}